\documentclass[aps,floatfix,lengthcheck,showpacs,amssymb,amsmath,amsfonts,twocolumn,nofootinbib,prd]{revtex4-1}
\usepackage{lineno}
\usepackage[utf8]{inputenc}
\usepackage{color}
\usepackage{graphicx}
\usepackage{float}
\usepackage{subfigure}
\usepackage{acronym}
\usepackage{multirow}
\usepackage{tikz}
\usetikzlibrary{shapes.geometric, arrows, calc, fit, positioning, chains, arrows.meta}
\usepackage{hyperref}
\hypersetup{
    colorlinks=true,
    linkcolor=blue,
    filecolor=magenta,      
    urlcolor=cyan,
}
\usepackage[toc, nonumberlist, acronym]{glossaries}
\glsdisablehyper
\usepackage{verbatim}
\usepackage[binary-units=true, alsoload=astro]{siunitx}
\usepackage{bm}
\usepackage{mathtools}
\usepackage{rotating}

\begin{document}

\newacronym{gw}{GW}{Gravitational wave}
\newacronym{usr}{USR}{unbounded Softmax replacement}
\newacronym{far}{FAR}{false-alarm rate}
\newacronym{bbh}{BBH}{binary black hole}
\newacronym{snr}{SNR}{signal-to-noise ratio}
\newacronym{psd}{PSD}{power spectral density}
\newacronym{fap}{FAP}{false-alarm probability}

\title[]{
From One to Many: A Deep Learning Coincident Gravitational-Wave Search
}

\author{
    Marlin B. Sch{\"a}fer$^{1,2}$,
    Alexander H. Nitz$^{1,2}$
}
\address{$^1$Max-Planck-Institut f{\"u}r Gravitationsphysik,
         Albert-Einstein-Institut, D-30167 Hannover, Germany}
\address{$^2$Leibniz Universit{\"a}t Hannover, D-30167 Hannover, Germany}

\begin{abstract}
Gravitational waves from the coalescence of compact-binary sources are now routinely observed by Earth bound detectors. The most sensitive search algorithms convolve many different pre-calculated gravitational waveforms with the detector data and look for coincident matches between different detectors. Machine learning is being explored as an alternative approach to building a search algorithm that has the prospect to reduce computational costs and target more complex signals. In this work we construct a two-detector search for gravitational waves from binary black hole mergers using neural networks trained on non-spinning binary black hole data from a single detector. The network is applied to the data from both observatories independently and we check for events coincident in time between the two. This enables the efficient analysis of large quantities of background data by time-shifting the independent detector data. We find that while for a single detector the network retains $91.5\%$ of the sensitivity matched filtering can achieve, this number drops to $83.9\%$ for two observatories. To enable the network to check for signal consistency in the detectors, we then construct a set of simple networks that operate directly on data from both detectors. We find that none of these simple two-detector networks are capable of improving the sensitivity over applying networks individually to the data from the detectors and searching for time coincidences.
\end{abstract}

\maketitle

\section{Introduction}
Gravitational waves (\acrshort{gw}s) are now routinely observed by the two Advanced LIGO detectors \cite{TheLIGOScientific:2014jea} and the Advanced Virgo detector \cite{TheVirgo:2014hva}. At the end of the last observing period, the KAGRA detector \cite{KAGRA:2018plz} joined the network and is expected to aid observations in the future. During three observing runs $\approx 90$ \acrshort{gw}s from compact binary sources have been identified, almost all of which are consistent with the merger of \acrfull{bbh} systems \cite{LIGOScientific:2018mvr, Abbott:2020gw2, Nitz:2021ogc, LIGOScientific:2021qlt, LIGOScientific:2021djp, Nitz:2021zwj}.

Many searches for \acrshort{gw}s from compact-binary coalescence use matched filtering to separate potential signals from the background detector noise \cite{Abbott:2020gw2, Messick:2016aqy, DalCanton:2020vpm, Adams:2015ulm}. Matched filtering is a technique that convolves a set of pre-calculated template waveforms, each representing a possible source with different component masses, spins, etc., with the detector's data and is known to be optimal for Gaussian noise \cite{Allen:2005fk}. A \acrfull{snr} time series is calculated for each template waveform; candidates are identified by a peak in the \acrshort{snr} time series that also passes data quality \cite{Nuttall:2015dqa, LIGOScientific:2016emj, LIGOScientific:2019hgc} checks. In a second step the candidate detections from one detector are cross-validated with the candidate detections from other detectors to further increase the significance of the reported events and rule out false positives \cite{Abbott:2020gw2, Usman:2015kfa, Sachdev:2019vvd}. For sources where the gravitational-wave signal is unknown or poorly modeled other search algorithms detect coincident excess power in different detectors and do not require a model \cite{Klimenko:2015ypf}.

Deep learning has started to be explored as an alternative approach to building an algorithm to detect \acrshort{gw}s \cite{George:2016hay, George:2017pmj, Gabbard:2017lja, Dreissigacker:2020xfr, Schafer:2020kor, Krastev:2020skk, Wei:2020ztw, Wei:2020sfz, Cuoco:2020ogp, Huerta:2021ybd}. It may potentially target signals which are currently challenging for matched filter search algorithms due to computational limitations \cite{Wei:2020ztw, Wei:2020xrl, Huerta:2020xyq}. The computational cost of these modeled searches scales with the number of templates required by the parameter space. Certain effects like higher-order modes \cite{Harry:2017weg}, precession \cite{Harry:2016ijz}, eccentricity \cite{Nitz:2019spj,Lenon:2021zac}, or the inclusion of sub-solar mass systems \cite{Nitz:2021mzz, Nitz:2021vqh} potentially require millions of templates and are thus computationally prohibitive to analyze. Deep learning may also be more sensitive when the noise is non-Gaussian \cite{Zevin:2016qwy, Essick:2020qpo, Wei:2020sfz}.

In our previous work \cite{Schaefer:2021trs} we explored the sensitivity of a simple neural network to non-spinning \acrshort{bbh} sources in Gaussian noise for a single detector. We tested how different training strategies influence the training procedure and the final efficiency of the network. Our results showed that under the given conditions the network can closely reproduce the sensitivity of matched filtering and that most efficient convergence is reached when a range of low \acrshort{snr} signals is provided throughout training.

Here we extend our previous work to two detectors. To do so, we use the same single detector network explored in \cite{Schaefer:2021trs} and apply it individually to the data from both observatories. This procedure produces a list of candidate events for each detector. We then search for coincident events between the two, where two events are assumed to be coincident if they are within the maximum time-of-flight difference between both detectors. We assume this difference to be \SI{0.1}{\second} since the networks are trained to be insensitive to variations on such scale.

The network uses the \acrfull{usr} modification we introduced in \cite{Schaefer:2021trs}. It outputs a single detector ranking statistic. Here we use it to construct a network ranking statistic. This network ranking statistic turns out to be the sum of the individual ranking statistics minus a correction factor.

The main advantage of this approach is the trivial computation of the search background which enables robust detection claims at comparable statistical significance ($<1$ per $100$ years) to existing production methodology. By applying time shifts larger than the time-of-flight difference between the detectors to the data from only one observatory, we can create large amounts of data which by construction cannot contain any astrophysical coincident candidates. By applying the time shifts to the single detector events rather than the input data directly, we can skip re-evaluating the entire test set and efficiently look for coincident events. This is a well established method that has already been successfully applied \cite{Abbott:2020gw2, Usman:2015kfa, Sachdev:2019vvd}. By this approach we can probe the search down to a false-alarm rate (\acrshort{far}) of 1 false-alarm per $\mathcal{O}\left(10^3\right)$ months. The \acrshort{far} estimates how often a candidate is produced by the search under the null hypothesis of no astrophysical candidates. Our \acrshort{far}-estimate is limited by the assigned hardware resources rather than the available data.

We compare this search to an equivalent matched filter search \cite{pycbc_version}. We find that the deep learning search still retains $92.4\%$ of the sensitivity of a two-detector matched filter search when the latter is restricted to using the timing difference between the detectors as the only means for determining coincident events. However, the matched filter search also extracts some information on the parameters of the signal. When we also require matching templates and the phase and amplitude of the triggered templates to be consistent between detectors \cite{Nitz:2017svb}, the machine learning search only retains $83.9\%$ of the sensitivity.

We then construct a single network that operates on the data from both detectors. The idea is that the network may then be able to learn, summarize, and cross-correlate signal characteristics between detectors. To do so, we remove the last layer of the original networks applied to the individual detectors and concatenate their output. Thereby the input data are compressed to a $128$ dimensional latent space. Dense layers are used to correlate the concatenated outputs and condense it into a single ranking statistic.

Using a single network complicates the background estimation, as time shifts between the detectors can in principle not be applied after evaluating the individual data streams. However, the two-detector network architecture is constructed such that the data from different detectors is analyzed by individual sub-networks, concatenated and processed by a third sub-network. This enables us to process the bulk of the data only once and apply time shifts to the individual detector sub-network outputs. To obtain the ranking statistic we are then only required to run the time-shifted data through the final, small sub-network.

We find that networks constructed this way are not able to improve the sensitivity over a time coincidence analysis of the single detector machine learning events. We test three different approaches to training these networks but none show any improvement.

\section{Coincident Search from Independent Single-Detector Networks}\label{sec:methods}
The algorithm explored in this section uses a network trained on data from a single detector and uses it to find coincidences in multiple detectors. It is one of the most simple extensions and has two advantages. First, networks trained on data from a single detector can be re-used which reduces requirements to computational resources. Second, the search background can be estimated using well established and efficient algorithms allowing for much higher confidence in candidate detections.

\subsection{Architecture}
We use the same network as in \cite{Schaefer:2021trs}, which is an adaptation of the network presented in \cite{Gabbard:2017lja}. It consists of 6 stacked convolutional layers followed by 3 dense layers. An overview of the architecture is given in \autoref{tab:network}.

\begin{table}[]
    \begin{ruledtabular}
    \caption{A detailed overview of the architecture for the single detector neural network. Rows are grouped by their influence on the shape of the data. The layers are to be read from left to right and top to bottom to construct the network.}\label{tab:network}
    \begin{tabular}{lrr}
        layer type & kernel size & output shape \\
        \hline
        Input + BatchNorm1d & & $2048\times 1$ \\
        Conv1D + ELU & 64 & $1985\times 8$ \\
        Conv1D & 32 & $1954\times 8$ \\
        MaxPool1D + ELU & 4 & $488\times 8$ \\
        Conv1D + ELU & 32 & $457\times 16$ \\
        Conv1D & 16 & $442\times 16$ \\
        MaxPool1D + ELU & 3 & $147\times 16$ \\
        Conv1D + ELU & 16 & $132\times 32$ \\
        Conv1D & 16 & $117\times 32$ \\
        MaxPool1D + ELU & 2 & $58\times 32$ \\
        Flatten & & $1856$ \\
        Dense + Dropout + ELU & & $64$ \\
        Dense + Dropout + ELU & & $64$ \\
        Dense + Softmax & & $2$
    \end{tabular}
    
\end{ruledtabular}
\end{table}

The last layer contains a Softmax activation function, which we remove during testing. In \cite{Schaefer:2021trs} we showed that this modification, which we called \acrfull{usr}, allows the network to be tested at lower \acrshort{far}s than otherwise possible.

The Softmax activation for the first output neuron is given by
\begin{equation}\label{eq:dx_to_p}
p \coloneqq {\text{Softmax}\left(\bm{x}\right)}_0 = \frac{1}{1+\exp\left(-\Delta x\right)},
\end{equation}
where $\bm{x} = \left( x_0, x_1 \right)$ is the network output before the activation function and $\Delta x = x_0 - x_1$. When $\Delta x$ is strongly positive, the denominator in \eqref{eq:dx_to_p} and thus the fraction numerically evaluates to $1$. This leads to problems when setting the threshold value to use to determine true positive detections \cite{Schaefer:2021trs}.

However, equation \eqref{eq:dx_to_p} is bijective and can be inverted
\begin{equation}\label{eq:p_to_dx}
-\Delta x = \log\left[\frac{1}{p} - 1\right].
\end{equation}
This quantity is monotonic and we can thus do statistics on $\Delta x$ directly, avoiding numerical instabilities while still using the Softmax activation during training.

\subsection{Data Sets and Training}\label{sec:training:data}
The input to the network is a time series of \SI{1}{\second} duration sampled at \SI{2048}{\hertz}. This allows for signals up to a frequency of \SI{1024}{\hertz} to be resolved which is sufficient for the considered parameter space.

The network is trained on signals from non-spinning \acrshort{bbh}s with component masses $m_1, m_2$ uniformly distributed from \SIrange{10}{50}{M_\odot}. We enforce $m_1\geq m_2$ and for each pair of masses uniformly draw $5$ coalescence phases $\phi_0\in\left[0,2\pi\right]$. The signals are generated with the waveform model \verb|SEOBNRv4_opt| \cite{Devine:2016ovp} (optimized version of \verb|SEOBNRv4| \cite{Bohe:2016gbl}) and scaled to varying optimal \acrshort{snr}s in the range $\left[5,15\right]$ during training. The time of merger is varied from \SIrange{0.6}{0.8}{\second} from the start of the input window to decrease the dependency of the network on the exact signal position. Each signal is whitened by the analytic model for the detector power spectral density (\acrshort{psd}) \verb|aLIGOZeroDetHighPower| \cite{lalsuite}. For further details on the training set please refer to \cite{Schaefer:2021trs}.

Notably, we do not vary the sky position, inclination or polarization during training. For a single detector, variations in these parameters can be fully expressed by changes in the distance, which is fixed by choosing a specific \acrshort{snr}, and the phase $\phi_0$. For a two detector setup this degeneracy is broken as a time-of-flight difference is introduced and the amplitudes and phases are correlated in the two detectors. However, our search algorithm is largely parameter agnostic. This means that its output does not depend on the amplitude or phase. Thus, we do not have information on whether or not the search responds to consistent signals. Finally, the time-of-flight difference is on the order of the variation of the merger time within the training set and can, therefore, not be resolved. In \autoref{sec:two-det-net} the network has access to data from both observatories and the data is adjusted accordingly.

All noise is Gaussian and simulated from the \verb|aLIGOZeroDetHighPower| \acrshort{psd} \cite{lalsuite}. We explicitly generate colored noise and whiten it afterwards. This in principle allows to extend our training to real noise.

The training set contains $200\,000$ noise samples, $100\,000$ of which are combined with $100\,000$ unique signals. The validation set\footnote{In out previous work \cite{Schaefer:2021trs} what we call validation set here was named efficiency set.} contains $400\,000$ noise samples and $10\,000$ unique signal samples, which we subsequently scale to \acrshort{snr}s $3, 6, 9, 12, 15, 18, 21, 24, 27\ \text{and } 30$. This set is used to calculate the \textit{efficiency} of the network at a fixed \acrfull{fap} of $10^{-4}$. The \acrshort{fap} is the fraction of discrete noise samples misclassified as signals. The efficiency is the fraction of discrete signal samples correctly classified as signals at a given \acrshort{fap}.

The test set contains a month of continuous simulated noise for each of the two detectors in Hanford and Livingston. We inject signals with parameters drawn from the distributions shown in \autoref{tab:injection} into both data streams. Injections are separated by a random time between \SIrange{16}{22}{\second}. To enable the networks to process this data, the continuous stream is sliced into $\approx 26$ million overlapping, correlated samples. Each sample is whitened individually by the analytic \acrshort{psd}.

We construct a second test set for background estimation. This set contains the same time domain noise as the first test set but no injections are performed. We pre-process this second data set in the same way we pre-process the first data set for the network to be able to process it.

The network is trained for $200$ epochs and we use the network with the highest average efficiency over all \acrshort{snr}s for the analysis carried out here. We use the Adam optimizer with a learning rate of $10^{-5}$, $\beta_1=0.9$, $\beta_2=0.999$ and $\epsilon=10^{-8}$ \cite{Kingma:2014ada}. We use a variant of the binary cross-entropy which was designed to stay finite as loss function
\begin{equation}
    L(\bm{y}_\text{t}, \bm{y}_\text{p}) = -\frac{1}{N_\text{b}}\sum_{i=1}^{N_\text{b}} \bm{y}_{\text{t},i}\cdot\log\left(\epsilon + (1 - 2 \epsilon) \bm{y}_{\text{p},i}\right),
\end{equation}
where $\bm{y}_\text{t}$ is ${\left(1,0\right)}^T$ for a signal-class sample and ${\left(0,1\right)}^T$ for a noise-class sample, $\bm{y}_\text{p}$ is the prediction of the network, $N_b=32$ is the mini-batch size, and $\epsilon=10^{-6}$.

We implemented the network using the high-level API Keras \cite{chollet2015keras} of TensorFlow version 2.3.0 \cite{tensorflow2015-whitepaper}.

\subsection{Single Detector Events}\label{sec:methods:single_det}
To apply the network to data of duration longer than the \SI{1}{\second} input of the network, we use a sliding window with step size \SI{0.1}{\second}. The contents of each window are whitened individually by the \acrshort{psd} model. At each step the network outputs a set of two numbers, the difference of which we use as our ranking statistic.

We apply the same network to the data from both detectors individually. We, thus, receive two output time series of ranking statistics. To determine notable events in the individual detectors we apply a threshold to both time series and cluster the resulting points above the threshold into events. A point exceeding the threshold is counted toward a cluster if it is within \SI{0.2}{\second} of the cluster boundaries. We choose a threshold on the \acrshort{usr} output of $-2.2$, which corresponds to a Softmax output of $0.1$.

The search algorithm produces a list of events, where an event is a tuple $\left(t, {\Delta x}\right)$. Each event is a time $t$ at which the network predicts a signal to be present with a ranking statistic ${\Delta x}$. The ranking statistic can be used to assign a significance to the event.

\subsection{Coincident Events}\label{sec:methods:coincs}
A signal will be present in the data of all detectors if it is of astrophyiscal origin. Its \acrshort{snr} in each detector depends on the location and orientation of the source. The number of false alarms can, thus, be reduced by requiring that the event is picked up by multiple detectors at similar times.

To quantify the significance of an event detected by more than one observatory, a combined ranking statistic is required. For simplicity we restrict our current analysis to two detectors. However, this approach is extendable to any number of detectors.

If the network was using the final Softmax activation during evaluation a combined ranking statistic would come straightforwardly from the interpretation of the output as a probability.
\begin{equation}\label{eq:combined_ranking_prob}
p_{H+L} = 1 - \left(1 - p_H\right)\left(1 - p_L\right)
\end{equation}

The 1-to-1 relation between $p$ and $\Delta x$ given in equation \eqref{eq:dx_to_p} can be inserted into \eqref{eq:combined_ranking_prob} to get
\begin{align}\label{eq:combined_ranking_ml}
-\Delta x_{H+L} = &-\Delta x_H - \Delta x_L \nonumber\\
                  & - \log\left[1 + e^{-\Delta x_H} + e^{-\Delta x_L}\right].
\end{align}
The combined ranking statistic is the sum of the single detector ranking statistics minus a correction term.

We consider an event in one detector to be coincident with another event in the other detector if the event times $t_i$ are within \SI{0.1}{\second} of each other. This time difference is chosen to be the maximum time resolution the networks can achieve due to the time variation in the training set.

We construct a list of coincidence events from the single detector list by the above condition. Each coincident event is assigned the combined ranking statistic \eqref{eq:combined_ranking_ml} and the time in the Hanford detector.

\subsection{Background Estimation}\label{sec:methods:background}
To estimate the \acrshort{far} at different ranking statistic values we evaluate the same noise used to search for signals but omit injecting the \acrshort{gw}s. This ensures that all events found in this data set are noise artifacts and are not influenced by close by injections.

We apply the network to the data and determine events as described in \autoref{sec:methods:single_det}. We obtain two lists of events and search for coincidences as detailed in \autoref{sec:methods:coincs}.

The lowest \acrshort{far} that can be probed is limited by the duration of the analyzed data. Our test set covers one month. The duration can be increased by shifting the data in one of the detectors by a time larger than the maximum time-of-flight duration between the detectors. Rather than shifting the data itself one may instead alter the event times returned by the search. This allows us to skip reanalyzing the full data for each time step and only requires us to look for coincidences between the events from one detector and the time shifted events from the second detector. Increasing the amount of background by applying time shifts is a well established method that has already been successfully applied in production searches \cite{Abbott:2020gw2, Usman:2015kfa, Sachdev:2019vvd}.

We choose a time shift of \SI{1024}{\second} and apply any possible integer multiple of this step size. We then search for coincidences in these events as detailed in \autoref{sec:methods:coincs}. This procedure increases our background to $\approx 2400\ \text{months} = 200$ years.

A list of \acrshort{far}s at different network ranking statistics is obtained by counting the number of events in the way described above with a larger ranking statistic.

\subsection{Sensitivity}
The sensitive volume of a search can be estimated by
\begin{equation}\label{eq:sens_vol}
    V\left(\mathcal{F}\right) \approx V\left(d_\text{max}\right)\frac{N_s\left(\mathcal{F}\right)}{N_\text{inj}},
\end{equation}
when it is derived on data containing injections which are distributed uniformly in volume \cite{Usman:2015kfa}. Here $\mathcal{F}$ is the \acrshort{far} at which the volume is being calculated, $d_\text{max}$ is the maximum distance of any injection, $V\left(d_\text{max}\right)$ is the volume of a sphere with radius $d_\text{max}$, $N_s\left(\mathcal{F}\right)$ is the number of signals detected with a \acrshort{far} $\leq\mathcal{F}$ and $N_\text{inj}$ is the total number of injected signals. We report the radius of a sphere with volume $V\left(\mathcal{F}\right)$ instead of the sensitive volume.

We analyze a month of simulated data from the two detectors Hanford and Livingston, assuming the \acrshort{psd} \verb|aLIGOZeroDetHighPower| \cite{lalsuite}. The data contains injections drawn from the distribution shown in \autoref{tab:injection}. We apply the network to the data from both detectors individually as described in \autoref{sec:methods:single_det}. The resulting single detector events are correlated and a list of coincident events is produced as detailed in \autoref{sec:methods:coincs}. We then pick out any events that are within \SI{0.3}{\second} of an injection. These events are called foreground events from here on out.

\begin{table}
    \caption{Distributions of the parameters used for the injections in the test set.}
    \label{tab:injection}
    \begin{ruledtabular}
    \begin{tabular}{lr}
        Parameter & Uniform distribution \\
        \hline
        Component masses & $m_1, m_2\in\ $\SI[parse-numbers=false]{\left(10, 50\right)}{M_\odot}\\
        Spins & 0\\
        Coalescence phase & $\Phi_0\in\left(0, 2\pi\right)$\\
        Polarization & $\Psi\in\left(0, 2\pi\right)$\\
        Inclination & $\cos{\iota}\in\left(-1, 1\right)$\\
        Declination & $\sin{\theta}\in\left(-1, 1\right)$\\
        Right ascension & $\varphi\in\left(-\pi, \pi\right)$\\
        Distance & \SI[parse-numbers=false]{d^2\in\left(500^2, 7000^2\right)}{{\mega\parsec}^2}
    \end{tabular}
    \end{ruledtabular}
\end{table}

To determine the search background, we evaluate the same month of noise used to find the foreground events. However, this data does not contain any injections. The networks return a list of single detector events, which are correlated and shifted in time to increase the effective duration of the analyzed data as detailed in \autoref{sec:methods:background}. The resulting coincident events are called background events from here on out.

We can then assign a \acrshort{far} to any foreground event. To do so we count the number of background events with a ranking statistic larger than the ranking statistic of the considered foreground event. This number is divided by the effective duration of the analyzed background to obtain a \acrshort{far}. The sensitive volume is then obtained from equation \eqref{eq:sens_vol} and converted to a distance. The sensitive distance as a function of the \acrshort{far} is obtained by evaluating the sensitive volume at the \acrshort{far}s of all foreground events.

\subsection{Matched Filtering}\label{sec:methods:matched_filter}
The template bank contains $598$ unique waveforms and is constructed such that no more than $3\%$ of the \acrshort{snr} of any signal is lost due to the discreteness of the bank. It covers the same mass range of \SIrange{10}{50}{M_\odot} as the training set of the networks and spins are set to $0$. The individual templates are generated using the waveform model \verb|IMRPhenomD| \cite{Husa:2015iqa, Khan:2015jqa} and placed stochastically.

To run the matched filter search we use the program \verb|pycbc_inspiral| \cite{pycbc_version}. It is setup to use a \acrshort{snr} threshold of $5$ in both detectors to create two sets of single detector triggers. These two sets are then checked for coincidence by two different approaches.

One approach handles the matched filter triggers analogous to the network single detector triggers, i.e.\ they are clustered and turned into single detector events as described in \autoref{sec:methods:single_det}. In this case the ranking statistic is the \acrshort{snr} returned by the best matching template. We then look for coincidences as described in \autoref{sec:methods:coincs} by requiring two events in different detectors to be separated by no more than \SI{0.1}{\second}. The combined ranking statistic in this case is given by
\begin{equation}\label{eq:combined_ranking_snr}
    \rho_{H+L}=\sqrt{\rho_H^2+\rho_L^2}.
\end{equation}
This disregards the information about the possible parameters obtained from the best matching template and only looks for time coincidence, i.e.\ no signal consistency is required.

The other approach leverages the signal information and checks for phase and amplitude correlation as well as requiring that the templates matching the data are consistent between detectors. In particular we utilize the combined ranking statistic given in equation (2) of \cite{Nitz:2017svb} and find coincidences as described therein.

\subsection{Evaluation and Comparison to Matched Filtering}\label{sec:net_sens}
\begin{figure}
    \centering
    \includegraphics[width=0.49\textwidth, trim=1cm 3.5cm 2cm 6cm, clip]{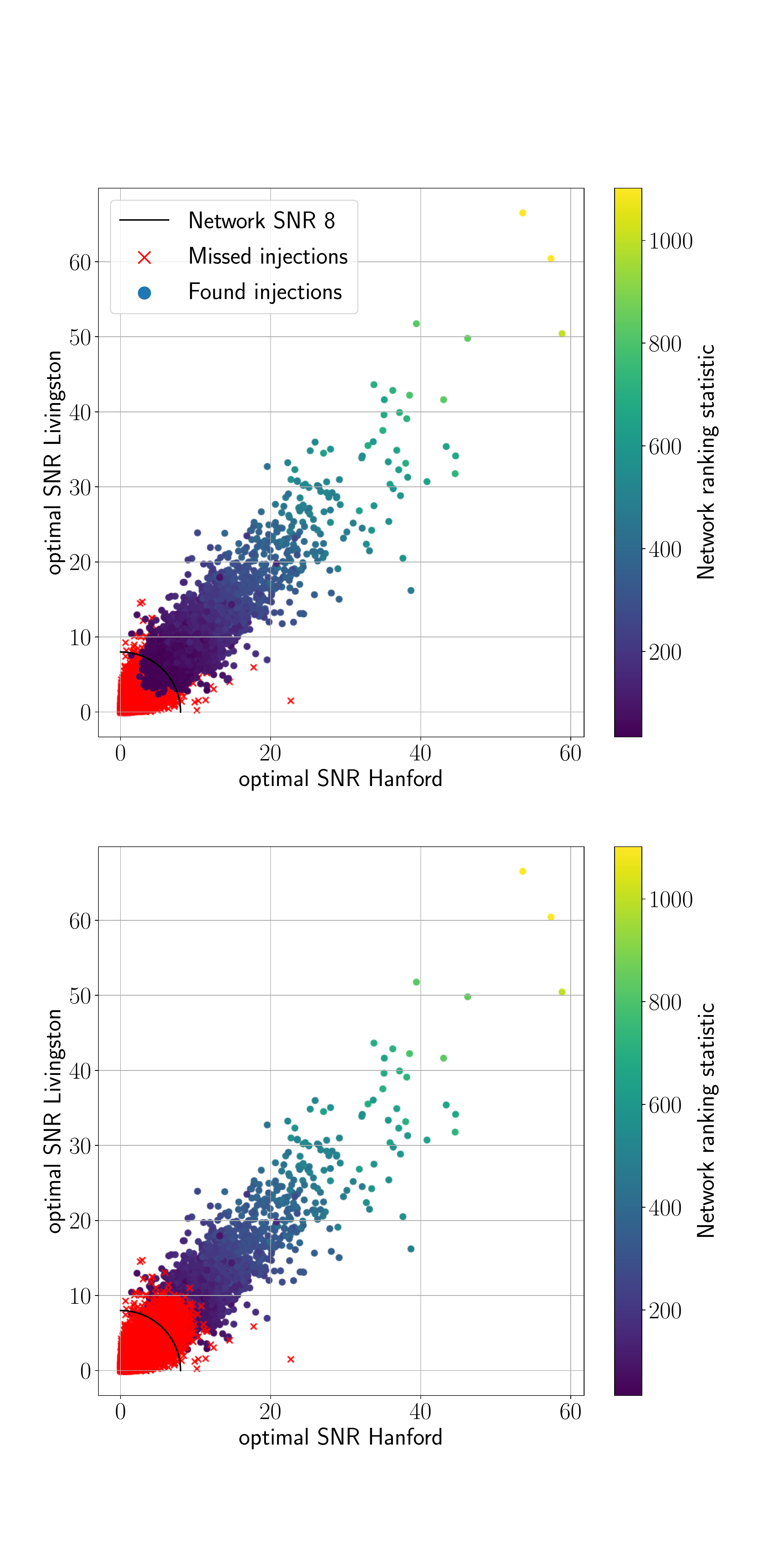}
    \caption{Found and missed injections from the test set as returned by the procedure discussed in \autoref{sec:methods}. The top panel overlays the missed injections by the found injections and the bottom panel reverses the order. The x- and y-axis show the optimal \acrshort{snr}s of the injections in the Hanford and Livingston detector, respectively. The color of found injections represents the combined ranking statistic as defined by equation \eqref{eq:combined_ranking_ml}. Missed injections are marked by a red cross. The black line indicates an optimal network \acrshort{snr} of $8$. The plot is generated at a \acrshort{far} of $1$ false alarm per month.}
    \label{fig:found_missed}
\end{figure}

In \autoref{fig:found_missed} we show the injections that were found and missed by the network coincident search at a \acrshort{far} of $1$ false alarm per month. The x-axis shows the optimal \acrshort{snr} of the injections in the Hanford detector and the y-axis shows the optimal \acrshort{snr} in the Livingston detector. The color indicates the network ranking statistic as calculated by equation \eqref{eq:combined_ranking_ml}. Missed injections are marked with a red cross. A network \acrshort{snr} of $8$ as calculated by equation \eqref{eq:combined_ranking_snr} is highlighted by the black line.

\autoref{fig:found_missed} shows that the combined ranking statistic \eqref{eq:combined_ranking_ml} is correlated with the network \acrshort{snr}. As the network \acrshort{snr} increases so does the combined ranking statistic. The loudest missed injection has a network \acrshort{snr} of $22.7$. However, the signal is most dominantly seen in the Hanford detector with a single detector \acrshort{snr} of $22.6$, whereas Livingston has an optimal \acrshort{snr} $<2$ due to the location of the source. Therefore, it is not surprising that the signal does not show up in both detectors and is missed by the coincidence search. When considering only the detector in which the signal is observable with lower \acrshort{snr}, the loudest missed signal has a optimal \acrshort{snr} of $9.2$ in that detector.

\begin{figure}
    \centering
    \includegraphics[width=0.49\textwidth, trim=1cm 0 4.2cm 2.5cm, clip]{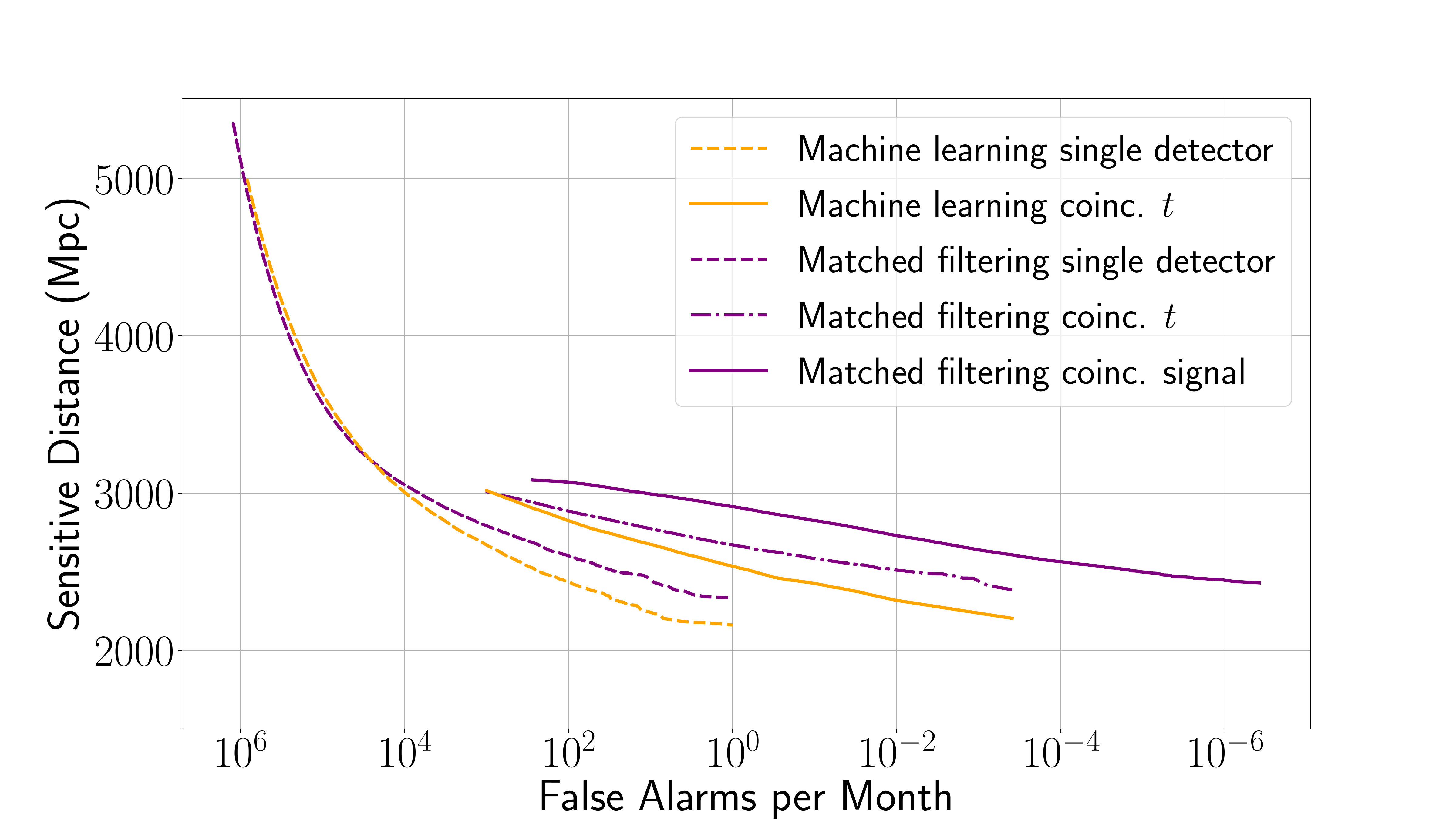}
    \caption{Shown are the sensitive distances of different search algorithms as a function of the \acrshort{far}. In orange we show the sensitivity curves of the machine learning based searches presented in \cite{Schaefer:2021trs} and this work. In purple we show sensitivity curves of an equivalent matched filter search. The dashed lines are derived on data only from a single detector. A label ''coinc. $t$'' refers to events being tested for coincidence based solely on the time difference of the events in the two detectors. The label ''coinc. signal'' means that the matched filter search also checked for signal consistency based on the time-, phase-, amplitude-difference, and intrinsic parameters in the two detectors. Sensitivities derived on data from more than one detector are truncated at a \acrshort{far} of $10^3$ per month due to an increasing number of true detections caused by random coincident events in the noise.}
    \label{fig:sens}
\end{figure}

In \autoref{fig:sens} we show the sensitive distance of different algorithms as a function of the \acrshort{far}. The orange lines show the sensitivity curves of the machine learning based algorithms whereas the purple lines show the sensitivities of a comparable matched filter search. The dashed lines show the sensitivity of the searches when only a single detector is considered. We compare those to a two-detector search where we require coincident detections in both detectors. The filled orange line and the dash-dotted purple line show the comparison between the machine learning and matched filter algorithms, respectively, when both impose the same coincidence condition. The filled purple line shows a more realistic application of matched filtering where the consistency of the time of arrival, the phase, the amplitude, as well as the parameters of the best matching template are required.

We find a significant improvement of up to $20\%$ at a given \acrshort{far} when the machine learning algorithm has access to data from both detectors compared to using only data from a single detector. Furthermore, we can probe \acrshort{far}s down to $\approx 4\times 10^{-4}$ false alarms per month without needing to increase the amount of evaluated data by applying time shifts between detectors as described in \autoref{sec:methods:background}. In principle this limit may be decreased even further and time shifts are only limited by the time-of-flight difference between the detectors. The large increase in the available background potentially greatly increases the statistical significance of any event.

The sensitivities of the machine learning search algorithms are compared to an equivalent matched filter search. For the single detector searches given by the dashed lines in \autoref{fig:sens} we find that the machine learning algorithm retains at least $91.5\%$ of the sensitivity at a fixed \acrshort{far} of the matched filter analogue. This corresponds to a maximum absolute separation of \SI{200}{\mega\parsec}. This difference in sensitivity is basically unchanged when data from two detectors is considered and both the machine learning as well as the matched filter search calculate coincidences only based on the timing in the different detectors. The corresponding curves in \autoref{fig:sens} are the filled orange and the dash-dotted purple line, respectively. In this case, the machine learning algorithm retains at least $92.4\%$ of the sensitivity of the time coincidence matched filter search which corresponds to an absolute separation of \SI{180}{\mega\parsec}.

However, matched filtering also carries information about the intrinsic parameters of the source, the relative phase, and the relative amplitudes in the two detectors. This information can be used to further constrain coincidences and improve the ranking statistic \cite{Nitz:2017svb} by testing for signal consistency. We compare the time coincidence machine learning search (filled, orange line in \autoref{fig:sens}) to this matched filter coincidence search utilizing signal consistency checks (filled, purple line in \autoref{fig:sens}). The machine learning search now only retains at least $83.9\%$ of the sensitivity in \acrshort{far} regions where both are defined. This corresponds to an absolute separation of \SI{430}{\mega\parsec}.

We truncate the sensitivity curve of any search that has access to data from both detectors in \autoref{fig:sens} at a \acrshort{far} of $10^3$ false alarms per month. This is done due to a large number of true positives at high \acrshort{far}s originating from random noise coincidences. This means that the search returns a coincident event that is caused by a particular noise realization which happens to coincide with an injection with an optimal \acrshort{snr} below the trigger threshold. Many of these injections should thus not be recoverable but are detected at high \acrshort{far} due to these noise fluctuations. At a \acrshort{far} of $10^3$ per month we expect less then $\mathcal{O}(10)$ of these false associations. Another reason to only compare the sensitivity at low \acrshort{far}s of the machine learning and the matched filtering based searches are the thresholds used to find triggers. The matched filter search uses a threshold of \acrshort{snr} $5$ whereas the machine learning search uses a threshold on the \acrshort{usr} ranking statistic of $-2.2$. Because there is no direct relation between these two statistics, we cannot guarantee that both thresholds correspond to similar signal strengths. It may be possible that one search excludes weak signals which are found by the other based on this difference in the threshold.

The sensitivity difference between machine learning and matched filtering stays constant between using data from a single detector and using data from two detectors when matched filtering may only check for time consistency between detection candidates from the two observatories. The performance difference increases when matched filtering also checks for signal consistency. It is, therefore, reasonable to believe that a multi detector machine learning search may be more sensitive when it too can check for signal consistency. This would either require the single detector network to output parameter estimates of the detected signal alongside a ranking statistic or a single network that uses the data from both detectors as input. In the following \autoref{sec:two-det-net} we explore the second hypothesis.

\section{Two Detector Network}\label{sec:two-det-net}
The deep learning algorithm presented in \autoref{sec:methods} is significantly less sensitive than the full matched filter analysis that takes signal consistency into account. On the other hand, when the deep learning algorithm is compared to the matched filter search where signal consistency is ignored, the difference in sensitivity is comparable to the difference in sensitivity for a single detector. This gives reason to believe that the difference in sensitivity compared to the full matched filter search could be reduced when the network may operate on the data from both detectors and consider coincidences itself.

\subsection{Architecture}
\begin{figure}
    \centering
    \includegraphics[width=0.48\textwidth]{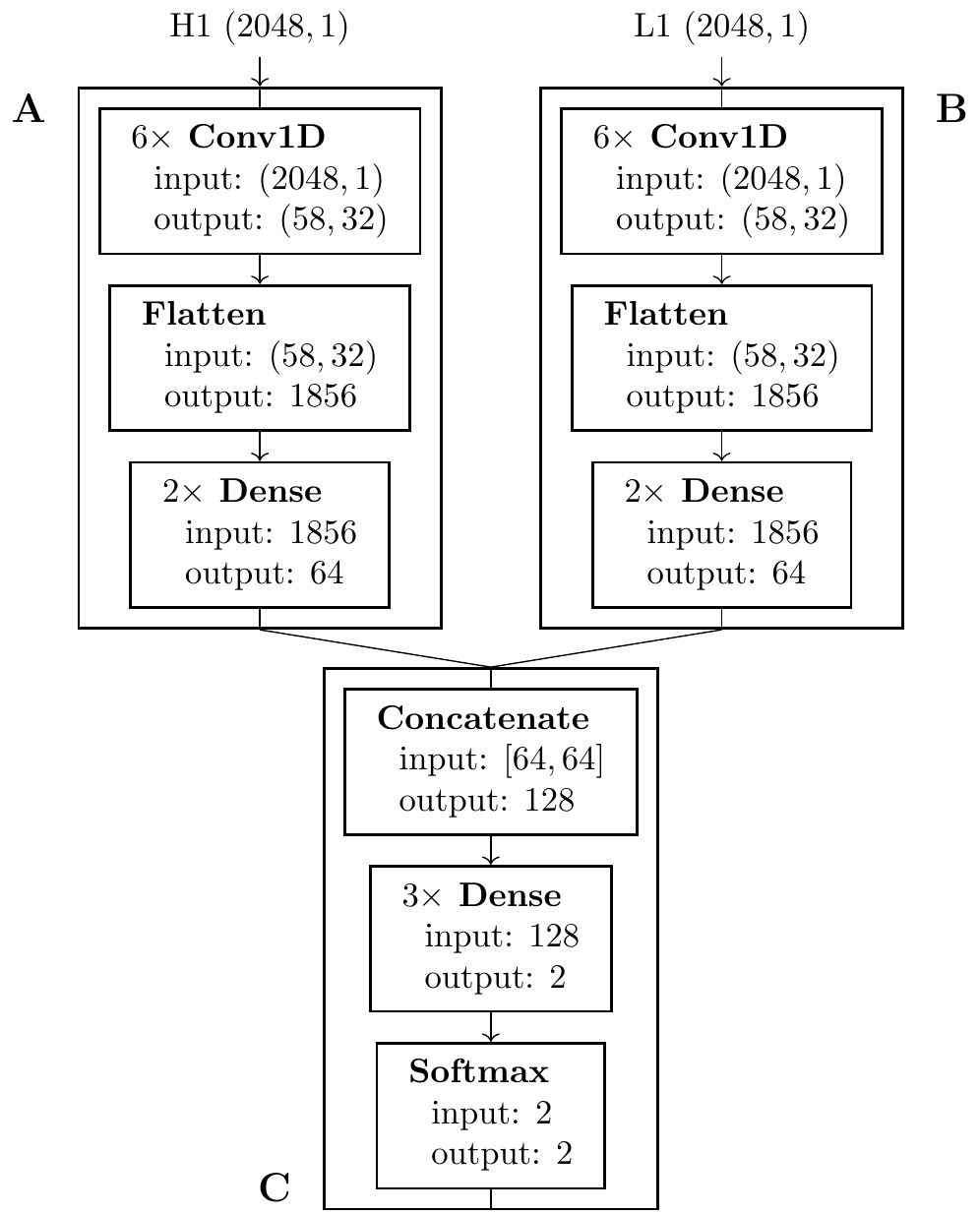}
    \caption{A high level overview of the two-detector architecture. The network consists of three sub-networks A, B, and C. A detailed description of the sub-networks A and B can be found in \autoref{tab:network} by removing the final row. The fully connected Dense layers contain $128$, $64$, and $2$ neurons in that order. All but the final Dense layer are equipped with an exponential linear unit (ELU) activation.}
    \label{fig:2det-net}
\end{figure}

We construct a network that uses data from both detectors while still retaining the ability to efficiently estimate a large background. The network from \autoref{sec:methods} is still applied to the data from the two detectors individually. However, the final layer is removed and the $64$ output-neurons from both networks are concatenated. We then add $3$ more fully connected layers to look for coincidences between the detectors. An overview of the network is shown in \autoref{fig:2det-net}.

The last layer from the single detector network is removed to create a large latent space. A matched filter search compresses the input data into the ranking statistic, the time of the merger, and the parameters of the best matching template. The intention is that $64$ neurons may be sufficient for a comparable compression and that the additional layers that operate on the concatenated outputs could perform a signal consistency analysis.

The sub-networks A and B in \autoref{fig:2det-net} are intended to act as encoders that reduce the $2048$ dimensional input into a latent space of dimension $64$. It may be interesting in the future to train these sub-networks initially as autoencoders \cite{Kramer:1991aaa} from which only the encoder is used for detection purposes afterwards. Autoencoders are neural networks which in the most simple form consist of an encoder network and a decoder network. The encoder network compresses the input to some lower dimensional latent representation whereas the decoder uses that lower dimensional representation to reconstruct the input. Other studies have already found that autoencoders have potential applications in \acrshort{gw} data analysis \cite{Shen:2019ohi, Gabbard:2019rde}.

\subsection{Data Sets and Training}
The network is trained on data similar to that presented in \autoref{sec:training:data}. However, the data is extended to two detectors and sources are uniformly distributed in the sky. The latter change is required due to the amplitude and phase correlations in the two detectors. We use the same number of noise and signal samples as in \autoref{sec:training:data}.

We utilize the pre-trained single detector network used in \autoref{sec:methods} in two different ways. In both cases the single detector parts of the two detector network (A and B in \autoref{fig:2det-net}) are initialized with the weights of the pre-trained model from \autoref{sec:methods}. However, for one of the two networks, these weights are then not optimized during training, leaving only the weights of the final fully connected layers (C in \autoref{fig:2det-net}) to be adjusted. This approach is known as transfer learning \cite{Weiss:2016aaa} and has been successfully applied for different problems \cite{Tan:2018, George:2018awu, Mesuga:2021qeq}. The second network optimizes the weights of the entire network. We also train a third network of the same architecture, where all parameters are initialized randomly and optimized during training.

The same optimizer settings and loss function described in \autoref{sec:training:data} are used to train all three networks for $300$ epochs. They are trained with a Softmax activation on the final layer, which is removed during evaluation. Each network is only trained once and the epoch with the highest efficiency on the validation set is chosen for further analysis.

\subsection{Coincident Events}
Because the networks output a single value when given the data from two detectors, we interpret that output as a coincidence ranking statistic at the corresponding time. We then perform the same clustering and thresholding described in \autoref{sec:methods:single_det} to obtain a list of coincident events.

\subsection{Background Estimation}
Determining the background of the two detector network is more challenging than for the single detector network from \autoref{sec:methods:background}, as there is no direct way of performing time shift in a computationally efficient way. One would, therefore, naively be limited by the duration of the analyzed data or would have to re-evaluate the entire month of test data multiple times. However, the network is designed in such a way that the data from both detectors are still analyzed individually and combined only at later stages. We evaluate the single detector data individually with the sub-networks A and B from \autoref{fig:2det-net} and store those outputs. We then permute the order of the outputs from sub-network B such that it corresponds to a time shift with respect to the output from sub-network A. Finally, sub-network C is applied to the concatenated data from sub-network A and B for many different time shifts. Since sub-network C is very simple and time shifts can be generated trivially this process generates $\mathcal{O}\left(1000\right)$ months of background within $<$ \SI{12}{\hour} on a NVIDIA RTX 2070 Super.

\subsection{Evaluation and Comparison to Matched Filtering}
\begin{figure}
    \centering
    \includegraphics[width=0.49\textwidth, trim=1cm 0 4cm 3cm, clip]{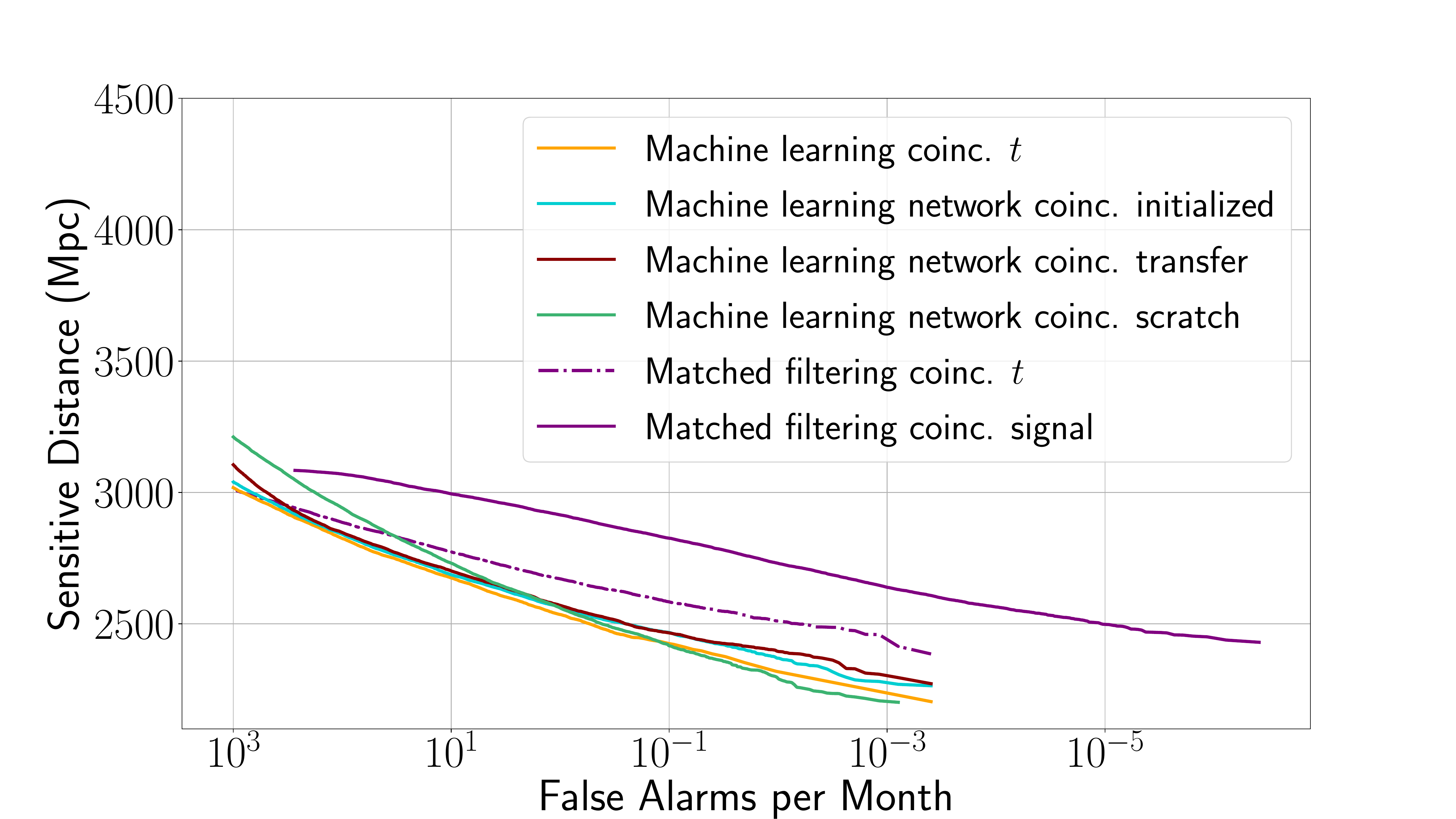}
    \caption{The sensitivity of different search algorithms as a function of the \acrshort{far}. All shown algorithms operate on the data from two detectors. The curves labeled ''Machine learning coinc.'' are neural network search algorithms that consider data from both detectors and an overview can be found in \autoref{fig:2det-net}. The network labeled ''initialized'' initializes the sub-networks A and B as shown in \autoref{fig:2det-net} from the single detector network used in \autoref{sec:net_sens} but optimizes them during the subsequent training. The network labeled ''transfer'' also initializes both sub-networks as the ''initialized'' network but freezes their weights. The network labeled ''scratch'' initializes all parameters of the network randomly. All other searches operate on the data from the individual detectors first and then search for coincident events. A label ''coinc. $t$'' refers to events being tested for coincidence based solely on the time difference of the events in the two detectors. The label ''coinc. signal'' means that the matched filter search also checked for signal consistency based on intrinsic parameters and the time-, phase-, and amplitude-difference in the two detectors. The curve labeled ''Machine learning coinc. $t$'' refers to the two-detector machine learning search analyzed in \autoref{sec:net_sens}. All sensitivities are truncated at a \acrshort{far} of $10^3$ per month due to a growing number of true positive detections caused by the coincidence of noise events.}
    \label{fig:dual_net_sens}
\end{figure}

\autoref{fig:dual_net_sens} shows the sensitive distance of the various networks as a function of the \acrshort{far} and compares them to the results presented in \autoref{sec:net_sens}. All curves are truncated at a \acrshort{far} of $10^3$ per month due to the large number of false associations described in \autoref{sec:net_sens}. The three networks utilizing the data from both detectors described in this section are labeled as ''Machine learning network coinc.''. The matched filter results are shown in purple, where the dash-dotted line considers only time coincidence and the filled line also takes the consistency of intrinsic source parameters, phase, and amplitude into account. The orange line corresponds to the network from \autoref{sec:methods}.

The networks described in this section were designed to be able to take signal consistency into account by reducing the input data to a large latent space. As such we were expecting sensitivities at low \acrshort{far}s to be larger than those obtained from time coincidence between single detector events produced by the single detector network.

However, we find that at low \acrshort{far}s all of the two detector networks are roughly as sensitive as the network tested in \autoref{sec:net_sens}. Therefore, they are still less sensitive than the matched filter equivalent and do not seem to take signal consistency into account. For high \acrshort{far}s, on the other hand, they are more sensitive. We suspect that the large time variation of the peak amplitude of \SI[parse-numbers=false]{\pm 0.1}{\second} may be responsible for this behavior. The networks are, thereby, trained to be insensitive to variations in timing of less then \SI{0.1}{\second}, which may produce phase and amplitude variations in a broad range.

\section{Conclusions}\label{sec:conclusions}
In this paper we have extended the single detector deep learning \acrshort{gw} search algorithm from \cite{Gabbard:2017lja, Schaefer:2021trs} to two detectors and compared it to an equivalent matched filter algorithm. We found that the most simple extension, applying the one detector network to the data from two detectors individually and searching for coincident events, retains $\approx 92\%$ of the sensitivity of matched filtering, when only the time consistency between detectors is required. This fraction drops to $\approx 84\%$ when signal consistency between detectors is also considered.

To operate on data from two observatories, we constructed a two detector ranking statistic for the machine learning search based on the single detector \acrshort{usr} ranking statistic proposed in \cite{Schaefer:2021trs}. This ranking statistic proved to be correlated with the network \acrshort{snr}.

We also highlighted the advantages of using a single detector network to construct a two detector search. First, the single detector network does not need to be re-trained to be applied to the second detector, if both have similar noise characteristics. Second, this approach enables an efficient background estimation by applying relative time shifts to the recovered single detector events. This allows to test the two detector search to almost arbitrarily low \acrshort{far}s at low computational expenses. This method has already proven to be effective and reliable in state-of-the-art classical search algorithms \cite{Abbott:2020gw2, Usman:2015kfa, Sachdev:2019vvd}.

Because using a single detector network restricts one to check for coincidences based solely on the timing difference, we tested a simple network that operates on data from both detectors directly. This allows the network in principle to construct internal signal representations which can be correlated between observatories. The network was constructed by removing the final layer of the single detector network, concatenating the outputs and adding a few fully connected layers to check for coincident events. The final fully connected layers, thus, receive $64$ latent variables for each detector that can be checked for coincidence.

This design of the two detector network allowed us to do efficient background estimation. By applying relative time shifts to the outputs of the individual detector sub-networks, only the final few fully connected layers need to be evaluated for all shifts. The bulk of the computation, namely evaluating the input data of the detectors, only needs to be done once.

The network architecture was trained in three different ways; randomly initialized parameters for the entire network, parameters of the sub-networks initialized from the single detector network, and parameters of the individual detector sub-networks fixed to the single detector parameters and optimizing only the final fully connected layers.

We found that all of these networks have very similar performance at low \acrshort{far}s. Neither of them performed substantially better than the initial network that looked for time coincident events between the single detector network outputs. It, therefore, seems as if the network architecture explored here is unable to learn any additional information about the signal. This may be caused by the allowed time-variance of \SI[parse-numbers=false]{\pm 0.1}{\second} for signals in the training set, which may limit the time resolution of the network and thus overshadow correlations in any other parameters. More sophisticated network architectures with higher time resolution may improve our findings. First promising steps have already been taken by \cite{Wei:2020ztw, Huerta:2020xyq}. Using an autoencoder to find a more meaningful latent representation of the input data may also be of use.

While the sensitivity was not improved by using a single network to process the data of two detectors, we still want to highlight that the method of determining the background may be of use for future networks.

Here we limited our research to \acrshort{gw}s from non-spinning binary black holes with signal duration $<$ \SI{1}{\second} and Gaussian noise. Any of these simplifications are desirable to be lifted. Especially considering real noise may increase the gap in sensitivity between the single detector and multi detector search algorithm, by vetoing glitches. While we considered only two detectors an extension to a larger network should be trivial and may follow studies such as \cite{Davies:2020tsx}.

\section{Acknowledgments}
We thank Ond\v{r}ej Zelenka, Frank Ohme, and Bernd Br\"{u}gmann for valuable discussions and their scientific input. We acknowledge the Max Planck Gesellschaft and the Atlas cluster computing team at Albert-Einstein Institut (AEI) Hannover for support.

\bibliography{bibliography}

\end{document}